\documentclass{PoS}

\def\integral{{\it INTEGRAL}}
\def\suzaku{{\it Suzaku}}
\def\xmm{{\it XMM-Newton}}

\def\chandra{{\it Chandra}}

\title{Broadband X-ray properties of absorbed AGN}

\ShortTitle{The X-ray view of type 2  AGN}

\author{\speaker{A. De Rosa}$^a$, F. Panessa$^{a}$, L. Bassani$^{b}$, A. Bazzano$^{a}$, A. J. Bird$^{c}$, R. Landi$^b$, A. Malizia$^b$, M. Molina$^d$, P. Ubertini$^a$\\
\llap{$^a$} INAF/IASF-Roma, Italy\\
\llap{$^b$} INAF/IASF-Bologna, Italy\\
\llap{$^c$} School of Physics and Astronomy, University of Southampton, UK\\
\llap{$^d$} INAF/IASF-Milano, Italy\\

E-mail: \email{alessandra.derosa@iasf-roma.inaf.it}}

\abstract{In this paper we report on the broadband X-ray properties of a complete sample of 33 absorbed Seyfert galaxies hard X-ray selected with \integral. 
The high quality broadband spectra obtained with  both \xmm, and \integral-IBIS data are well reproduced with an absorbed primary emission with a high energy cutoff and its scattered fraction below 2-3 keV, plus the Compton reflection features.
A high energy cut-off is found in 30\% of the sample, with an average value below 150 keV.
The diagnostic plot N$_H$ $vs$ F$_{obs}$(2--10 keV)/F(20--100 keV) allowed the isolation of the Compton thick objects, and may represent a useful tool for future hard X-ray observations of newly discovered AGN.
We are unable to associate the reflection components with the absorbing gas as a torus, a more complex  scenario being necessary. In the Compton thin sources, a fraction (but not all) of the Fe K line needs to be produced in a gas possibly associated with the optical Broad Line Region, responsible also for the absorption.  We still need a Compton thick medium (not intercepting the line of sight) likely associated to a torus, which contributes to the Fe line intensity and produces the observed reflection continuum above 10 keV.
The so-called Iwasawa-Taniguchi effect can not be confirmed with our data.
Finally, the comparison with a sample of unobscured AGN shows that, type 1 and type 2 (once corrected for absorption) Seyfert  are characterized by the same nuclear/accretion properties (luminosity, bolometric luminosity, Eddington ratio), supporting the ''unified'' view. }

\FullConference{The Extreme and Variable High Energy Sky - extremesky2011,\\
		September 19-23, 2011\\
		Chia Laguna (Cagliari), Italy}

\begin{document}

\section{Introduction}

The widely accepted Unification Model (UM) for active galactic
nuclei \cite{antonucci93} is strongly supported by spectropolarimetric
observations of hidden broad-line regions (HBLRs) in several type 2 AGN, and by the  
 X-ray observations, showing that Seyfert 2  have absorption columns exceeding the Galactic ones \cite{cappi06}. Despite this, and in view of  several pieces of evidence coming from recent observations of different classes of AGN, the UM is still being debated.
In particular, the association of the Compton thick (N$_{H}\gtrsim \sigma_{T}^{-1} \backsim$ 10$^{24}$ cm$^{-2}$) obscuring medium with a  uniform torus at intermediate scale between the  BLR and NLR, has been questioned by the presence of N$_{H}$ variability on timescales of hours, requiring an X-ray absorber no larger than the BLR \cite{risaliti07}.  These absorbing media are also responsible for the reflected features in the X-ray range: a Compton hump above 10 keV and a Fe line at 6.4-6.9 keV \cite{perola02}.
This kind of study has been performed in a statistical way with a variety of different samples over a broadband energy range: BeppoSAX \cite{risaliti02}, \integral\ and \xmm\ \cite{derosa11,derosa08}, \suzaku\ and Swift-BAT (both type 1 and type 2 AGN \cite{eguchi09,winter09,fukazawa11}. 
The 4th INTEGRAL/IBIS catalogue \cite{bird10} includes more than 700 hard X-ray sources detected
over a large part of the sky above 20 keV  and with positional accuracies ranging between 0.2 and 5.5 arcmin, depending on the source significance.
To obtain the best quality broadband high energy data, the \integral/AGN team has undertaken a programme of  follow-up observations of hard X-ray selected AGN with \xmm\ \cite{molina08,derosa08,derosa11}
In this paper we report on the broadband spectral properties of 33 type 2 AGN included in the  hard X-ray selected AGN complete sample as  defined by \cite{malizia09} optically classified as type 1.8-2.
The main goals of this study, widely presented and discussed in \cite{derosa11}, are (1) to measure
the spectral properties of obscured AGN: the intrinsic continuum shape  together with the absorption distribution and (2) to
accurately constrain the strength of the reflection components. The line properties tell us about the obscuring/reflecting medium while the Compton continuum is an important yet uncertain parameter (like the high energy cut-off) in population
synthesis models of the CXB \cite{gilli07}.

\section{Spectral analysis}
\label{spectra}

We fitted simultaneously the soft and hard X-ray spectra available with \integral/IBIS and  \xmm\ in the 0.3-100 keV energy range. 
To take into account possible flux variation of the source between the non simultaneous \xmm\ and \integral\ data, we left  a cross-calibration constant, between IBIS and pn, free to vary during the fit procedure.
Possible miscalibration  between could mimic or hide the presence of a Compton reflection component above 10 keV; this issue is widely discussed in \cite{derosa11}.

For  the sake of clarity and to ease the comparison with the published results on the known objects, we considered a baseline model (BLM) which is commonly used to fit the broadband X-ray spectra (0.3--100 ekV) of absorbed AGN \cite{comastri10,ueda07,risaliti02,derosa08}. The BLM is composed of:
1) a primary continuum extending up to the hard X-ray energy domain  modelled by a cutoff power--law having photon index $\Gamma$ and high energy roll-over E$_{c}$;
2) a soft X-ray scattered component modelled by a power--law  having photon index  $\Gamma_{s}$ assumed to be equal to  $\Gamma$ and a scattering fraction f$_{sc}$;
3) a Compton reflection continuum above 10 keV with the inclination angle $\theta$ of the reflector fixed to 45 deg, and solar abundance, plus the Fe K$\alpha$ emission line that is reproduced with a narrow (FWHM$<$1000 km s$^{-1}$) gaussian profile; broad line cases will be discussed in Sect. \ref{refl_sec}
All the  spectral components are absorbed by the Galactic column density along the line of sight  N$_{H}^{Gal}$, while the primary power--law is also absorbed by a second cold medium with column density N$_{H}$.
%
%
The broadband spectra of IGR sources and the best fit parameters for the whole sample are reported in \cite{derosa11}

\section{Results and Discussions}
\subsection{The primary absorbed and scattered continuum}

\begin{figure}
\centering
\includegraphics[width=0.3\linewidth]{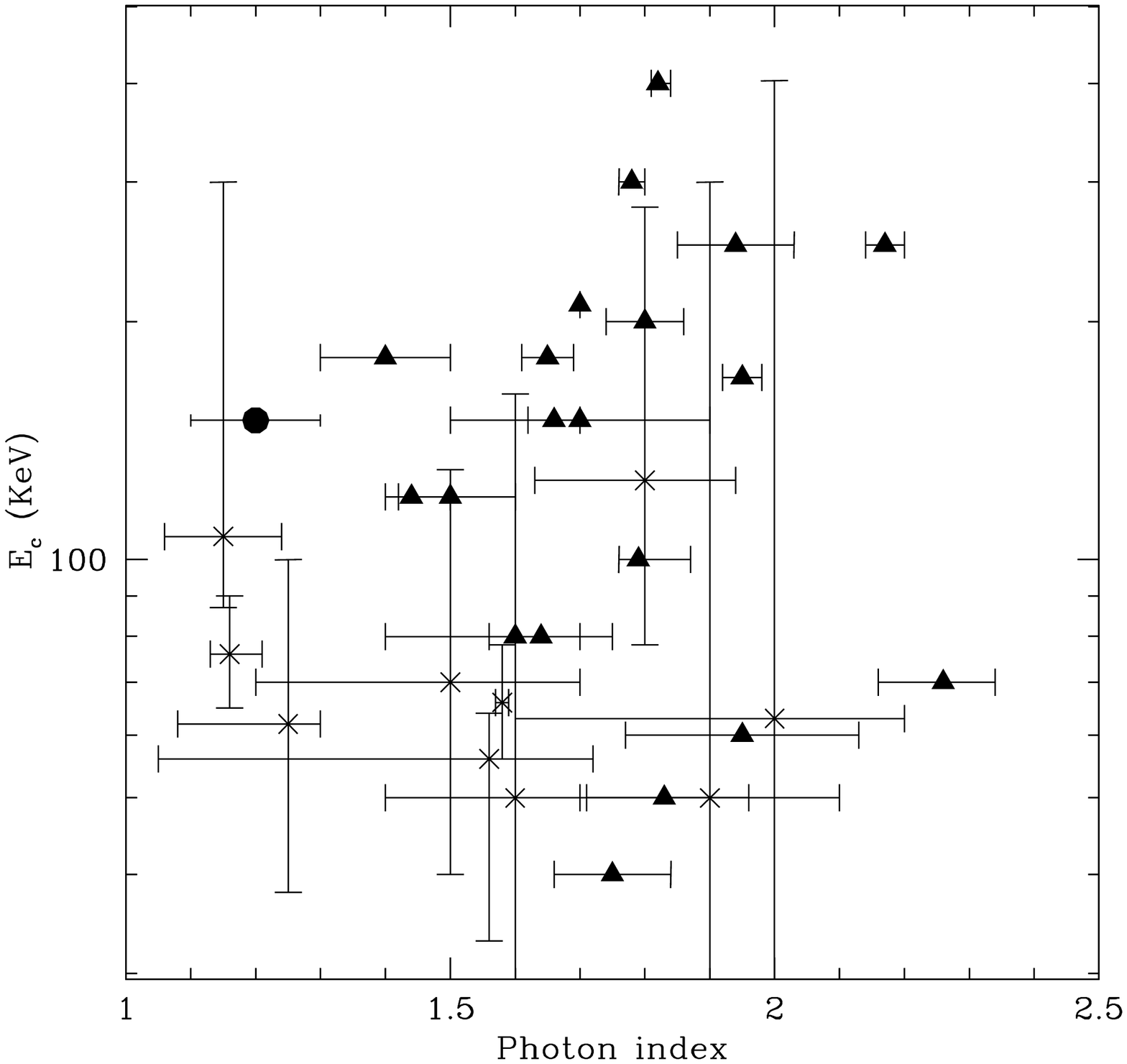}
\includegraphics[width=0.3\linewidth]{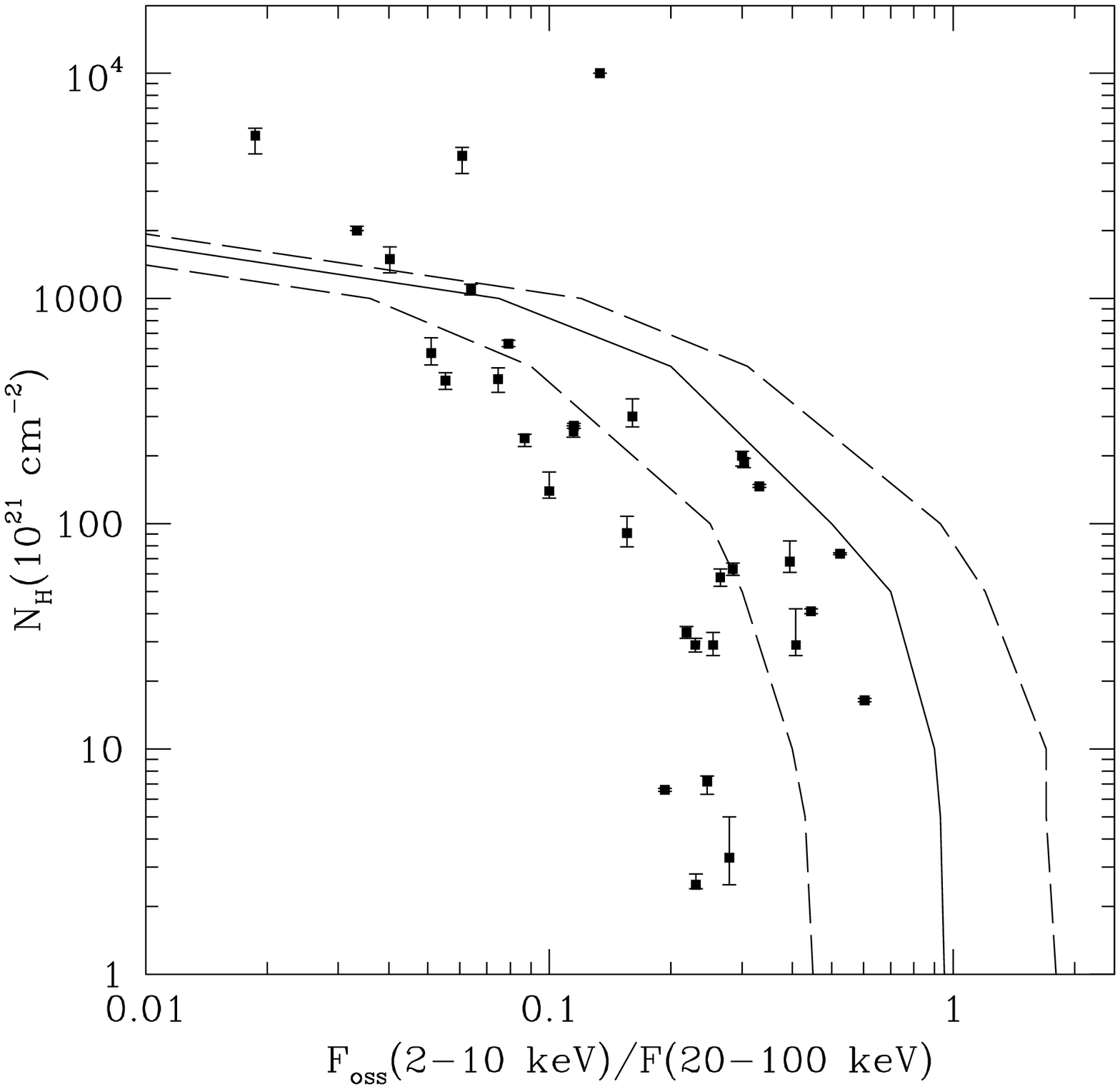}
\includegraphics[width=0.3\linewidth]{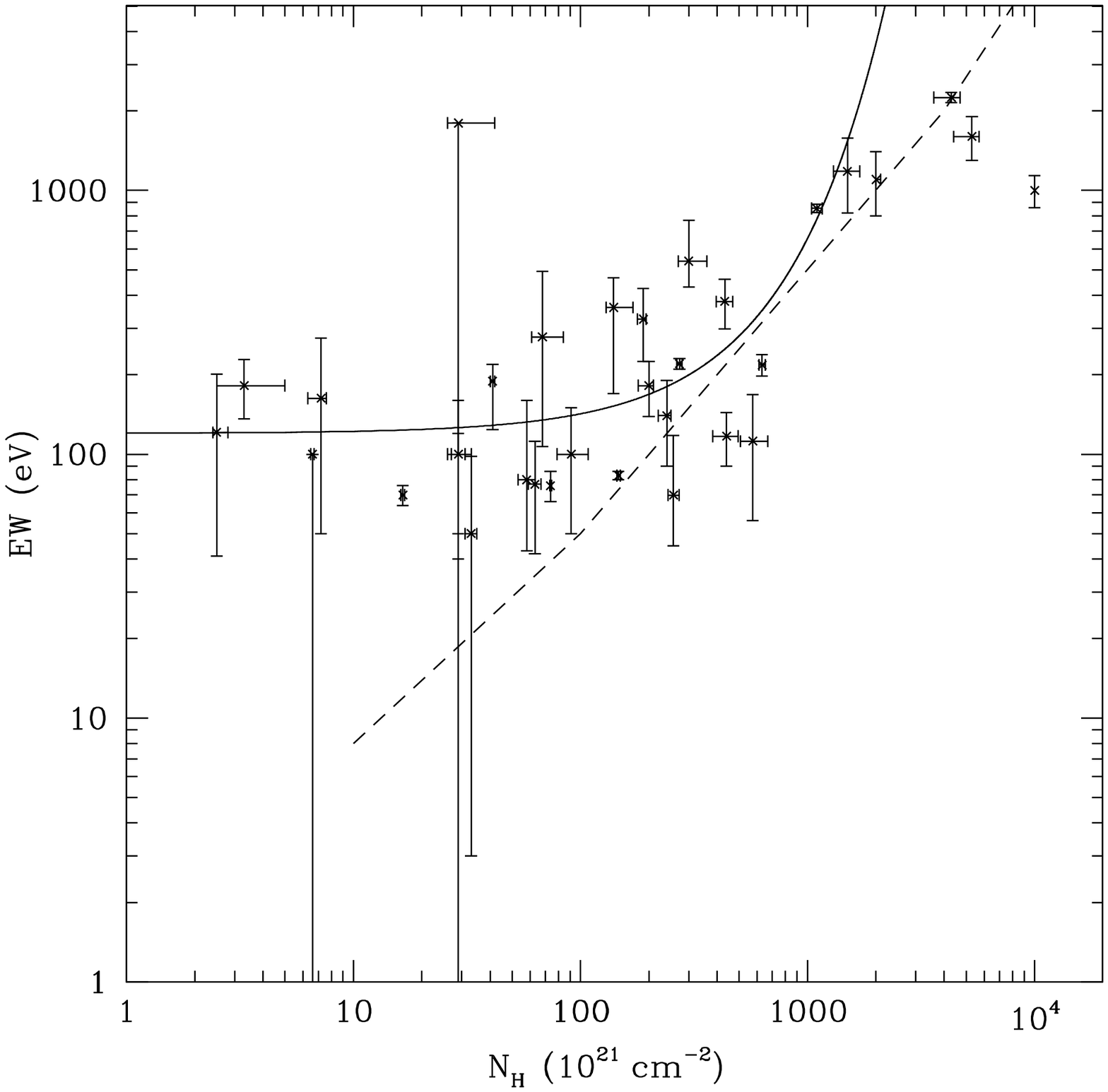}
\caption{ \textit{Left panel}: High-energy cutoff vs photon index for the whole sample.\textit{Central panel}: The column density distribution as a function of the softness ratio F$_{oss}$(2--10 keV)/F(20--100 keV). \textit{Right panel}: Fe K$\alpha$ EW  as a function of N$_{H}$. See text for details}
\label{fig1}
\end{figure}

The BLM  reproduces quite well all the spectra analysed since the $\chi^{2}$/dof ranges between 0.77-1.12.
The mean value of the photon index and its standard deviation for the whole sample is $\Gamma$=1.68, $\sigma$=0.30. This value is in very good agreement with those found in previous analyses performed in a limited energy range (2--10 keV) in both type 1 and type 2 AGN (e.g, \cite{singh11,bianchi09}).  
We measured the high energy cutoff in 10 out of 33 sources (i.e. 30 per cent of the sample), while only an upper limit at 300 keV has been found in IGR J16351-5806. In the remaining sources we obtained lower limits to E$_{c}$. The average value of the ten measured high energy cutoffs is E$_c$=73 keV  ($\sigma$=25 keV).
In Fig. \ref{fig1} (left panel) we plot E$_{c}$ as a function of $\Gamma$. All the measured values are below 150 keV, while lower limits (filled triangles) are well grouped below 300 keV. These values are consistent with those observed in the Seyfert 1 by BeppoSAX \cite{perola02,derosa07}. We do not find any evidence of correlation between $\Gamma$ and E$_c$,  the correlation coefficient being r=0.1.
Our analysis strongly suggests that the high energy rollover is an ubiquitous property of Seyfert galaxies, and that its value  ranges in the same intervals for type 2 and type 1 Seyferts \cite{molina08}. 

We found a value of $f_{sc}$ of a  few per cent in all sources (with the exception of IGR J14515-5542 and IGR J16351-5806  see  discussion in \cite{derosa11}). This value is in good agreement with that found in different samples of obscured AGN, observed with  \xmm\ and \chandra\ \cite{bianchi06}. No evidence of correlation between $f_{sc}$  and relative reflection R has been detected in our sample.

The average value of the absorption column density is 23.15 ($\sigma$=0.89). Excluding the six Compton thick sources we obtain an average value ($\sigma$) of 22.85 (0.67). 
In Fig. \ref{fig1}, central panel, we plot N$_{H}$ as a function of the ratio of the observed 2--10 keV flux $vs$ the 20--100 keV flux. This diagnostic plot was proposed by \cite{malizia07} in an attempt to isolate peculiar objects among  a sample of 34 \integral\ AGN observed with Swift-XRT.
The plot shows a clear anticorrelation between N$_{H}$ and the softness ratio: this ratio becomes lower as the column density increases.
The lines define a region obtained by assuming an absorbed power-law with high energy cutoff, where we have used  the average values of $\Gamma$ and E$_c$, with errors, as found in our sample.
In this diagnostic plot, Compton thick AGN are well isolated, all six Compton thick objects of our sample are located above the line.

\subsection{The reflection features: Fe K line and Compton hump}
\label{refl_sec}
We have  measured the reflection fraction (R) in 22 sources (average value R=1.63 with $\sigma$=1.15), while for 8 objects we only found upper limits.
With the sole exception of IGR J17513-2011, a neutral Fe line is detected in all the sources of the sample (marginally significant for IGR J00040+7020, 95\%).
The average value of the equivalent width (EW), calculated with respect to the total direct+reflected continuum
is 430 eV (with a somewhat high standard deviation of 564 eV),  while, excluding the Compton thick objects, it becomes 196 eV ($\sigma$=176 eV). It is worth noting that the fit of our BLM produces a degeneracy between the reflection fraction R and the photon index $\Gamma$: a  higher $\Gamma$  produces a higher value of R.
 Another point to stress is that the soft and high energy data (below and above 10 keV) are non-simultaneous. This point is discussed in \cite{derosa11}.
In the right panel of Fig. \ref{fig1} we show the Fe K$\alpha$ EW  as a function of N$_{H}$. 
Below N$_{H}$=10$^{23.5}$cm$^{-2}$ the EW trend is almost flat, with a best fit constant value of EW$_{0}$=120$\pm$16 eV; this value is in a very good agreement with that found in other X-ray sample of Compton thin  type 2 \cite{fukazawa11}  as well as type 1 AGN \cite{bianchi09} observed with  \suzaku, and \xmm, respectively.
However, above N$_{H}$=10$^{23.5}$cm$^{-2}$ the behaviour is different, and a clear correlation between N$_{H}$ and EW is evident, as expected from reflection from a Compton thick torus \cite{awaki91}. 
The dashed line in that figure represents the EW values as obtained  from transmission through a uniform slab of medium with solar abundances subtending 4$\pi$ to a continuum source of photon index $\Gamma$ = 2.0 \cite{leahy93}.  Large EW above 10$^{23.5-24}$cm$^{-2}$ could be originated in a Compton thick torus, also responsible for the observed absorption. On the other hand, the large EW  observed in Compton thin AGN (N$_{H}$ below 10$^{23.5}$cm$^{-2}$) are far above the dashed line, suggesting that the line is due to reflection from a material different from the absorber.
The evidence that the same Fe K$\alpha$ EW ($\sim$120 eV) is found  for Compton thin and type 1 AGN, suggests that the origin of this component is in the same medium, e.g. the BLR.
The solid line in  Fig. \ref{fig1}  represents the function EW(N$_{H}$)=EW$_{0}$exp($\sigma_{Fe} N_H$), with $\sigma_{Fe} $ being the photoelectric cross-section at 6.4 keV and EW$_{0}$=120 eV.
This trend well reproduces the Fe EW below 10$^{23.5}$cm$^{-2}$, and represents a scenario in which the Fe line is generated by reflection off the BLR seen along an unobscured line-of-sight, while the underlying continuum is obscured by matter with a column density $N_{H}$;  the total effect will be a larger value of EW increasing  N$_{H}$. 
It is worth noting that in 4 sources of our sample a broad Fe line is required to fit the broadband data. 
The intrinsic width of these  lines is completely compatible with an origin within the BLR.
This scenario is supported by the recent observations of rapid variation of the absorbing column density in both Compton thin and thick AGN. 
However, we stress that a Compton thin gas, also covering 4$\pi$ solid angle to the source, cannot create a Fe EW of 50-150 eV \cite{yaqoob01} as observed in our sample below N$_{H}$=10$^{23.5}$cm$^{-2}$; then, a Compton thick medium like a torus (not intercepting the line of sight in Compton thin or type 1 AGN), should exist.

Our conclusion is that in heavily obscured Seyferts (N$_{H}$ above 10$^{23.5-24}$cm$^{-2}$), the Fe line is produced in a torus also responsible for the absorption;  for Compton thin sources (N$_{H}$ below 10$^{23.5}$cm$^{-2}$), as well as for type 1 AGN, a high fraction of the Fe K line (but not the total line photons) has to be produced in a gas  located closer to the black hole than to the torus. This region can be associated with the BLR that could also be responsible for  the observed moderate absorption.
We also investigated the X-ray Baldwin/''Iwasawa-Taniguchi'' effect \cite{baldwin77} indicating the existence of an  anticorrelation between the 2--10 keV luminosity corrected for absorption and the EW of the Fe K line. 
A  linear fit  to the data  of the form log(EW)=5.3$\pm4.8$ - (0.07$\pm$0.11)log(L$_{corr}$(2--10 keV)), indicates no evidence of a correlation (r=0.109).
The X-ray Baldwin effect is not confirmed in our sample; although, we note that using the average EW in each luminosity bin, the correlation increases to r=0.2.  Considering the hard X-ray luminosity in the 20--100 keV energy range, the correlation coefficient increases in this case to r=0.3, still indicating a lack of correlation.  
It is worth noting that the luminosity range we are exploring is rather narrow (log(L)=41.5-44.5), and this could make it difficult to measure any correlation as observed by \cite{bianchi09} over a wider luminosity range.

%
%
%


\subsection{Type 1 $vs$ type 2 accretion properties}
\label{ty1_ty2}

\begin{figure}
\centering
\includegraphics[width=0.3\linewidth]{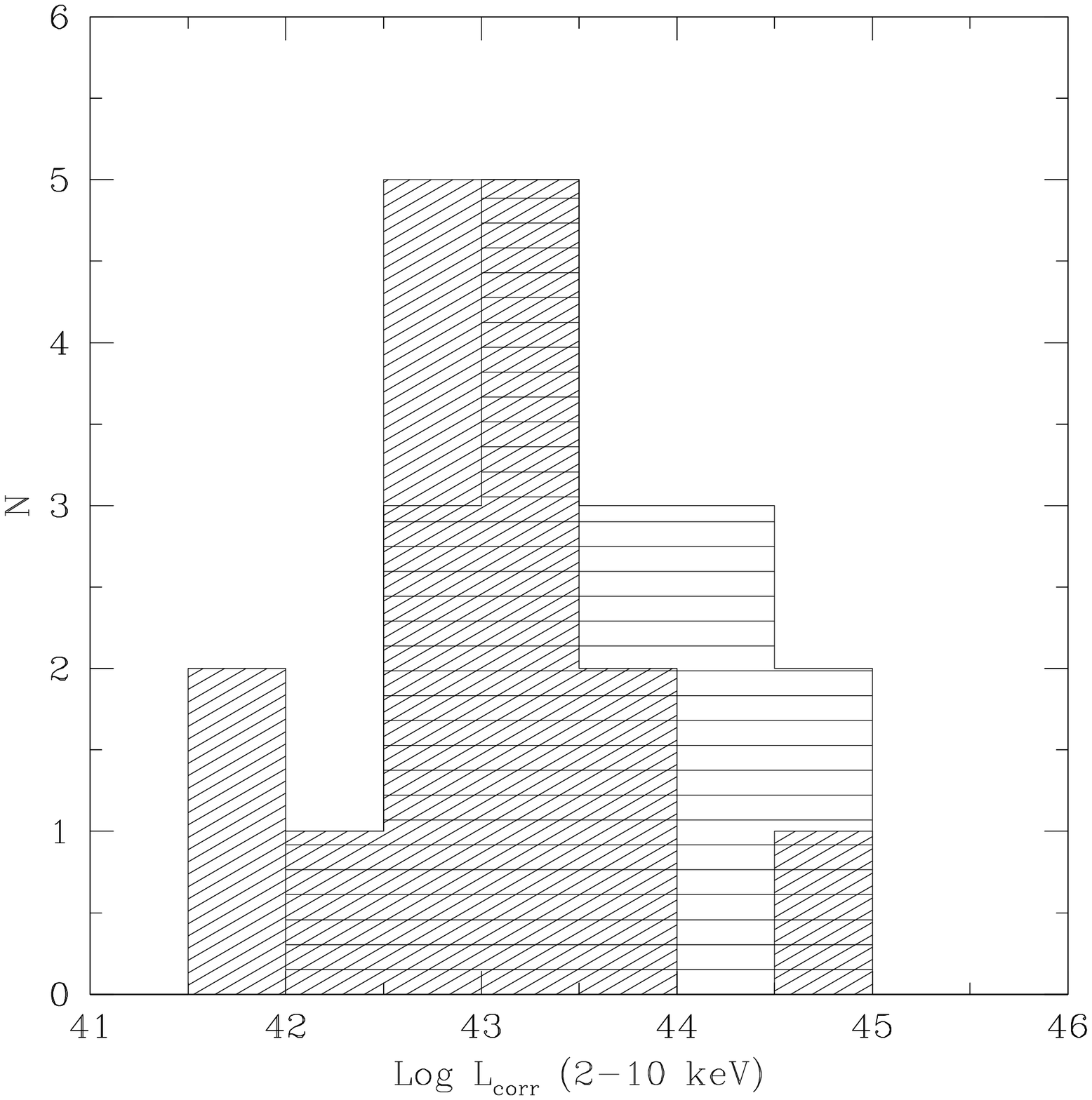}
\includegraphics[width=0.3\linewidth]{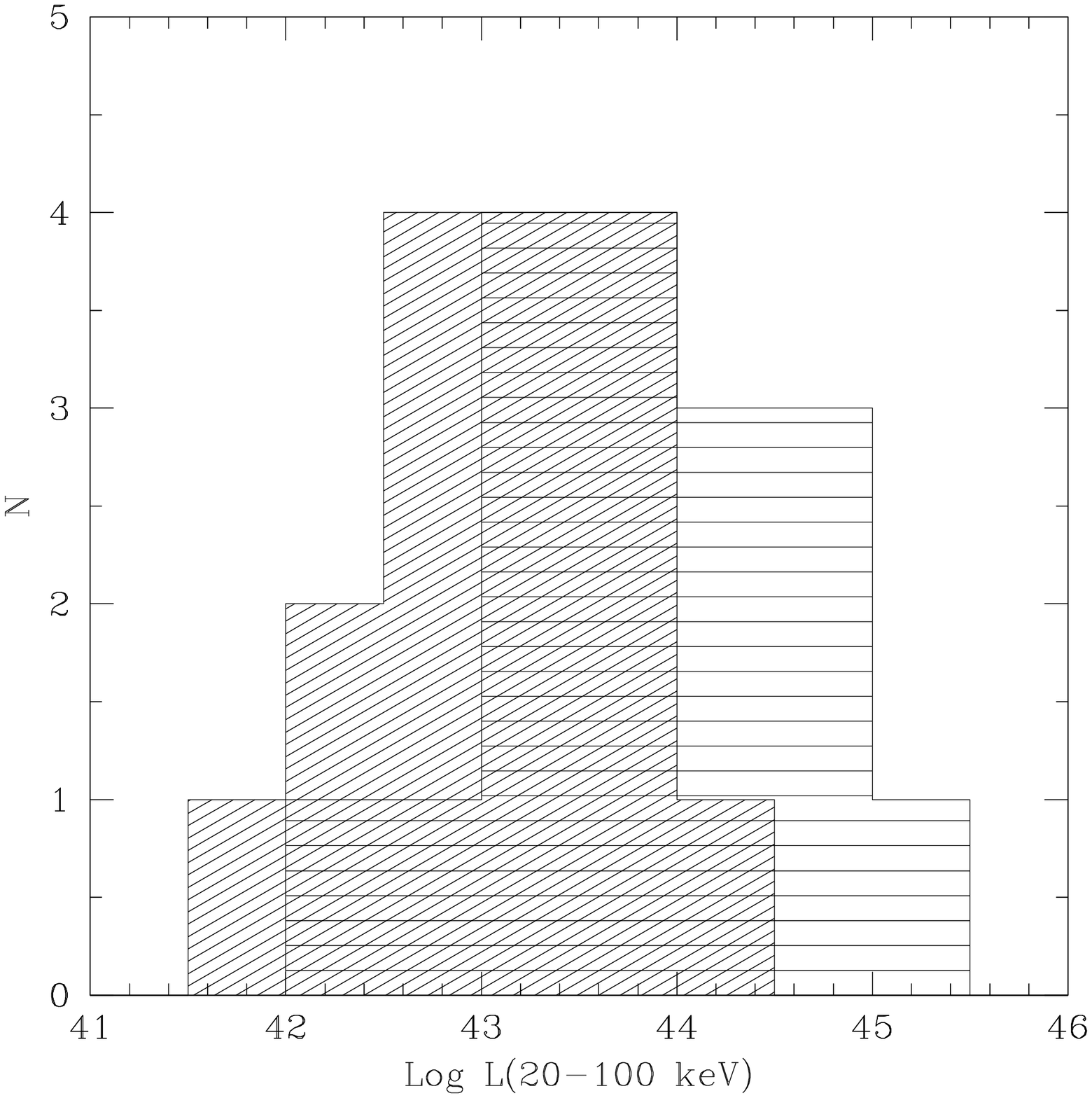}
\includegraphics[width=0.3\linewidth]{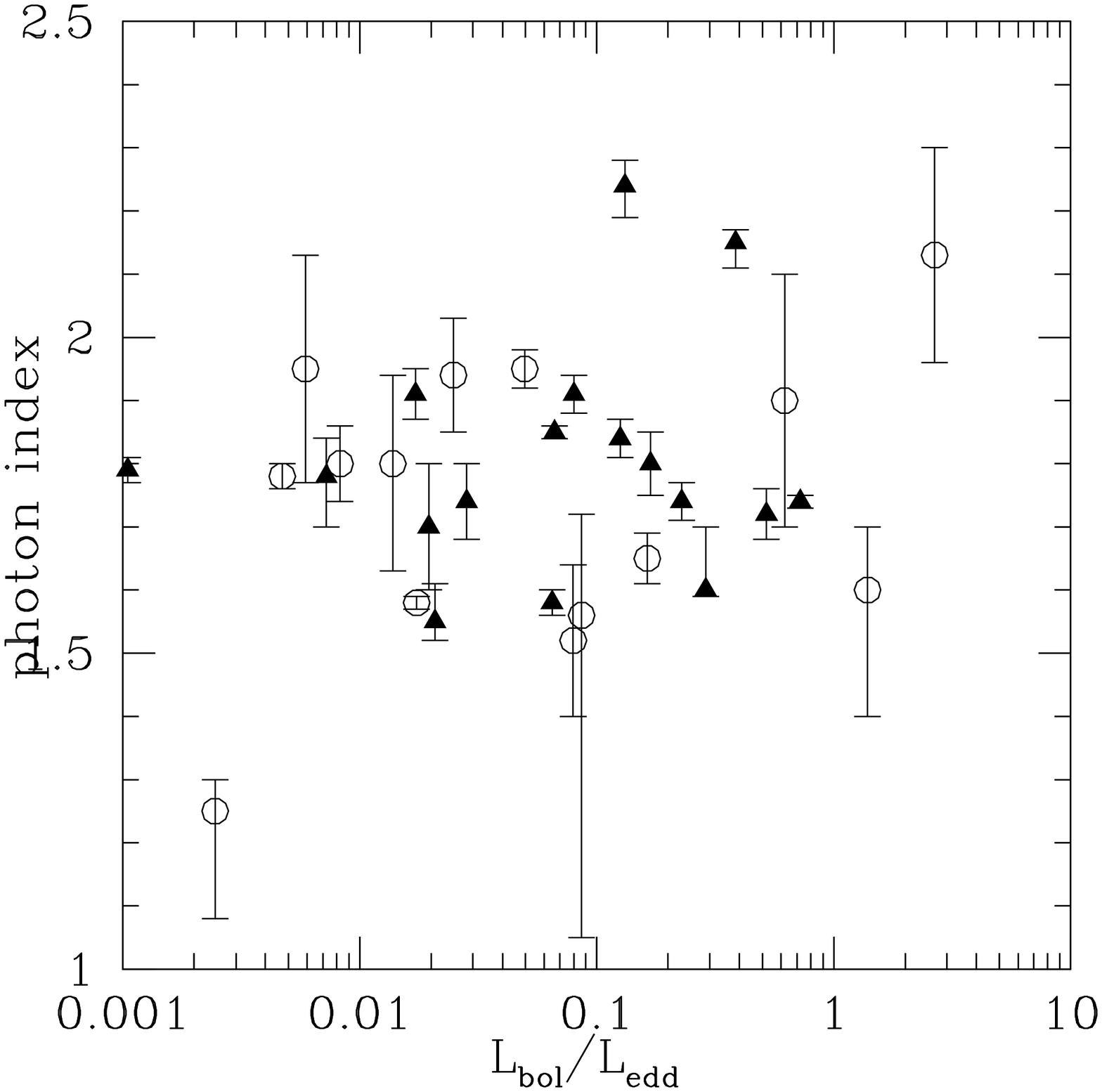}
\caption{\textit{Left and Central panels}: Luminosity distribution (corrected for absorption) in (2--10 keV) and 20--100 keV for the sub-sample of type 1 and type 2 AGN with measured masses. Type 1 are shown as horizontal large spaced  line, type 2 are diagonal small spaced lines. \textit{Right panel}: $\Gamma$ as a function of the ratio L$_{bol}$/L$_{edd}$  for the sub-sample of type 1 and type 2 AGN with measured masses. Filled triangles are type 1 while open circles represent type 2 AGN.}
\label{fig2}
\end{figure}

From the samples of hard X-ray selected obscured (this work) and unobscured \cite{molina08} AGN, 
we collected  all available published mass measurements obtained with reverberation mapping techniques, maser, X-ray variability or M-$\sigma$ relation. 
A sub-sample of 33 (17 type 1+16 type 2) AGN with known masses has been obtained. This sub-sample can be considered as representative of the whole one (see discussion in \cite{derosa11}).
We used this hard X-ray selected sample to draw  a statistical comparison between type 1 and type 2 Seyfert of (1) the absorption corrected 2--10 keV and 20--100 keV luminosity, L$_{corr}$(2--10 keV) and L(20--100 keV); (2) the bolometric luminosity, L$_{bol}$; and (3) the Eddington ratio, L$_{bol}$/L$_{Edd}$, with the main goal of testing the prediction of the UM for AGN. 
The distribution of the corrected 2--10 keV luminosity (left panel of Fig. \ref{fig2}),  has averaged value of 43.52 ($\sigma$=0.71) and 43.06 (0.72)  for type 1 and type 2, respectively. Using the  Kolmogorov-Smirnov test we do not find a significant difference in the distributions, having a probability  P=0.111 to be drawn from the same parent population.
This result is also confirmed in the higher, unbiased for absorption, energy range (20--100 keV) (see distribution in the central panel of Fig. \ref{fig2}), with a KS test probability of 0.137.
To be conservative and fully convinced to have properly corrected  the 2--10 keV luminosities for absorption, we excluded the  Compton thick sources from this analysis and, even in this case,  the KS test probability  confirms the similarity of the two populations (P=0.165).

For all sources in the sub-sample we calculated the bolometric luminosity  derived from the 2--10 keV luminosities, adopting the luminosity-dependent bolometric correction given in  \cite{marconi04}. 
The L$_{bol}$ average value is 44.91 ($\sigma$=0.91) and 44.33 (0.90) for type 1 and type 2, respectively. A KS test gives a probability P=0.111 that the distributions belong to the same parent population, suggesting, even in this case, the similarity between the two classes of AGN.
In the right panel of Fig. \ref{fig2} we plot $\Gamma$ versus the Eddington ratio for our sub-sample.
The average value of the ratio L$_{bol}$/L$_{edd}$  is 0.17 ($\sigma$=0.20)  and 0.32 (0.72) for type 1 and type 2 AGN, respectively. Although the dispersion is huge, we note that, as for the luminosity, no evidence of separation in the accretion rate distribution is evident between absorbed and unabsorbed AGN (KP probability P=0.292).
These  results concerning luminosities is also reflected in the spectral properties. In fact a comparison between the primary photon index in type 1 and type 2 class does not show any evidence for a separation being the average value of $\Gamma$ equal to 1.80 ($\sigma$=0.18) and 1.75 ($\sigma$=0.26) for type 1 and type 2 respectively. A KS test  in this case gives a probability P=0.862.

These same conclusions have  recently been reached in \cite{singh11} through the X-ray analysis of an optical selected sample of type 1 and type 2 Seyfert. 
The value of the L$_{bol}$/L$_{edd}$ ratio in our sub-sample of type 1 and type 2 AGN  range in 0.001 - 3, extending to higher values with respect to a optical selected sample \cite{singh11}
We are aware that this result could be affected by two issues: mass estimation and bolometric corresction, both are discussed in \cite{derosa11}
 Finally, we also stress that, using the whole hard X-ray selected sample of  41 type 1 + 33 type 2 , we find that L$_{corr}$(2--10 keV) and L(20--100 keV) have a  small  probability  (P=0.05 and 0.02, respectively) to be drawn from the same parent population. This effect is likely to be due to a bias in the sample selection through high lunimosity AGN.

\end{document}